\documentclass[11pt]{article}
\usepackage{moriond,epsfig}

\def\be{\begin{equation}}
\def\ee{\end{equation}}
\def\bea{\begin{eqnarray}}
\def\eea{\end{eqnarray}}

\begin{document}
\vspace*{4cm}
\title{LHC data and the proton strangeness}

\author{ N.P. Hartland }

\address{Tait Institute, School of Physics and Astronomy,\\ University of Edinburgh, EH9 3JZ, UK.}

\maketitle
\abstracts{
The LHC has already provided many relevant measurements for the determination
of parton distribution functions (PDFs). Measurements of the W and Z lepton distributions
are of interest for flavor separation and in particular for
the determination of the relatively poorly constrained strange
quark distribution. In this contribution we shall discuss the
computational developments that allow for the efficient inclusion of LHC data into the
NNPDF framework consistently at NLO for all observables, and we study the constraints of the LHC W and Z data on the strangeness content
of the proton.}
There have been a number of experimental measurements of direct relevance to PDF determination performed by LHC collaborations. Measurements of inclusive jet and dijet cross sections\cite{CMS:2011ab,Chatrchyan:2011qta,Aad:2010ad} and electroweak vector boson production\cite{Aad:2011dm,Aaij:2012vn,Chatrchyan:2011jz}  provide information on PDFs in previously unexplored kinematical regions. While the importance of including LHC data in future determinations is clear, of particular interest is the potential impact of these data sets upon collider only fits, where existing determinations tend to be poorly constrained. This necessitates the inclusion into PDF determinations low energy data with potential contamination from nuclear corrections or higher twist effects.

However, including collider measurements into a PDF fit on a large scale requires substantial computational resources. NNPDF parton sets are fitted by genetic algorithm minimisation over a large number of generations,\cite{Ball:2010de} therefore the NNPDF methodology requires a fast method of computing collider observables. Although LHC data was previously included in NNPDF2.2\cite{Ball:2011gg} by a reweighting method, the constraining power of the LHC dataset makes adding a large quantity of data in this manner impractical. We shall here describe a fast convolution method that has been developed to enable the inclusion of new hadronic data in NNPDF fits, before going on to discuss some preliminary results from fits including LHC data (NNPDF2.3 preliminary). There has been particular interest in the usefulness of the recent ATLAS measurements of W/Z production\cite{Aad:2011dm} in providing information on the strange content of the proton.\cite{Aad:2012sb} We shall discuss here some preliminary results on the proton strangeness fraction using the updated fit.

A number of tools are available for the computation of hadronic observables that allow for a straightforward variation of the input PDF \emph{a posteriori}, a prerequisite for
utility in parton fitting. 
In particular, the FastNLO\cite{Kluge:2006xs} and APPLGrid\cite{arXiv:0911.2985} projects provide software which is well suited for use in fitting. The principle of these projects is to
store the required perturbative coefficients for a process as weights upon an interpolating grid in $x$ and $Q$ space. The convolution required to calculate the observable is then reduced
to a simple product, the PDF in the product may be straightforwardly varied along with the chosen value of $\alpha_S$. For example, to compute a hadronic cross section in the APPLGrid framework,  the following calculation is performed,
\begin{equation}
\label{eq:applconv}
\sigma = \sum_p \sum_{l=0}^{N_{\mathrm{sub}}} \sum_{\alpha,\beta}^{N_x} \sum_{\tau}^{N_{Q}}
W_{\alpha\beta\tau}^{(p)(l)} \, \left( \frac{\alpha_s\left(Q^2_{\tau}\right)}{2\pi}\right)^{p}
F^{(l)}\left(x_{\alpha}, x_{\beta},  Q^2_{\tau}\right),
\end{equation}
where the indices $\alpha,\beta$ run over points in the $x$-space grid. $\tau$ runs over points in $Q^2$, $p$ denotes the perturbative order of the contribution, and $l$ denotes the specific parton level subprocess. The $W$ table contains the values of the Monte Carlo weights for a particular subprocess point, and the $F^{(l)}$ are the incoming subprocess parton densities constructed as a combination of PDFs as appropriate for the process in consideration. This method of computing observables is fast, but substantial speed improvements can be gained by combining PDF evolution with this procedure. For a set of flavour basis PDFs $f$, we write a general subprocess density as,
\be F^{(l)}\left(x_{\alpha}, x_{\beta},  Q^2_{\tau}\right) =  \sum_{i,j}^{13} C^{(l)}_{ij}  \left( f_i(x_{\alpha},Q^2_\tau)f_j(x_{\beta},Q^2_\tau) \right).  \ee
Where $i,j$ denote the PDF flavour, and the $C^{(l)}_{ij}$ are coefficients specifying how the subprocess density $l$ is to be built. The evolution of the initial state PDF to the required scale $Q^2_{\tau}$ can be performed in an analogous fashion to the convolution in Eqn \ref{eq:applconv} by evaluating the matrix of DGLAP evolution kernels upon an interpolation grid as per the FastKernel method.\cite{Ball:2010de} Obtaining the evolved PDF is reduced once again to a product,
\be f_i(x_{\alpha},Q^2_\tau) =  \sum_j^{13}R_{ij}N_j(x_{\alpha},Q_\tau^2) =\sum_{j}^{13} \sum_{\gamma}^{N_x}  \sum_{k}^{N_{\mathrm{pdf}}} R_{ij}E^\tau_{\alpha\gamma j k}N^0_k(x_\gamma).\ee 
Where the $N$ are PDFs in a suitable evolution basis that diagonalises the matrix of DGLAP evolution kernels. The matrix $E^\tau_{j k}$ holds the values of the DGLAP evolution kernel $\Gamma_{jk}\left(x,Q_0^2,Q_\tau^2 \right)$ convoluted with the interpolating basis functions as in reference \cite{Ball:2010de}, and the matrix $R$ is the rotation matrix from the evolution to the flavour basis. Here we adopt the notation  $N^0$ for the $N_{\mathrm{pdf}}$ light evolution basis PDFs parameterised at the initial fitting scale $Q^2_0$. It is now simple to construct the subprocess density using these matrices,
\be F^{(l)}\left(x_{\alpha}, x_{\beta},  Q^2_{\tau}\right) =   \sum_{i,j}^{13} \sum_{k,l}^{N_{\mathrm{pdf}}}  C^{(l)}_{ij}A^\tau_{\alpha\gamma i k}A^\tau_{\beta\delta j l}N^0_k(x_\gamma)N^0_l(x_\delta),
\quad\quad A^\tau_{\alpha\gamma i k} =  \sum_j^{13} R_{ij}E^\tau_{\alpha\gamma j k}.\ee
With the PDF evolution factorized, the computation in Eqn \ref{eq:applconv} is now reduced to a much simpler form particularly suited to a fitting application,
\be \sigma =  \sum_{i,j}^{N_{\mathrm{pdf}}} \sum_{\alpha,\beta}^{N_x} \widetilde{W}_{\alpha\beta i j} N_i^0(x_\alpha)N_j^0(x_\beta),\ee
where 
\be \widetilde{W}_{\alpha\beta i j} = \sum_p \sum_{l=0}^{N_{\mathrm{sub}}}\sum_{k,l}^{13} \sum_{\gamma,\delta}^{N_x} \sum_{\tau}^{N_{Q}}
W_{\gamma\delta\tau}^{(p)(l)} \, \left( \frac{\alpha_s\left(Q^2_{\tau}\right)}{2\pi}\right)^{p} C^{(l)}_{kl}A^\tau_{\gamma\alpha k i}A^\tau_{\delta\beta l j}, \ee
is the weight matrix containing all the values that may be precomputed and stored prior to a PDF fit. The calculation of a hadronic observable is then simply a matter of
 a sum of products over a grid in $x$-space, and the now reduced flavour basis of $N_{\mathrm{pdf}}$ light PDFs. Through this method we are able to reproduce the results of the original
APPLGrid/FastNLO calculation at the same level of precision and with a substantial improvement in speed.

Using this technique, we can now present results on the strangeness fraction with recent preliminary NNPDF fits including LHC data. $R_s (x,Q^2)=(s+\bar{s})/(\bar{u} + \bar{d})$ has been determined from a number of NNPDF fits at NLO to different datasets. Firstly a fit (here denoted NNPDF2.3 prelim) to the full NNPDF2.1 dataset with the addition of ATLAS $35$ $\mathrm{pb}^{-1}$ inclusive jet measurements,\cite{Aad:2010ad} ATLAS $35$ $ \mathrm{pb}^{-1}$ W and Z rapidity distributions,\cite{Aad:2011dm} and CMS $840$ $\mathrm{pb}^{-1}$ W electron asymmetry data.\cite{CMSW840} Secondly, a fit exclusively to the NNPDF2.3 collider data subset (NNPDF2.3 Collider), and finally, a fit to the HERA-I combined dataset\cite{HERA1comb} and ATLAS W/Z measurement only (NNPDF2.3 HERA+ATLASWZ). For comparison, the value of $R_s (x,Q^2)$ determined by the NNPDF2.1\cite{Ball:2011mu} fit is also provided.

\begin{figure}[ht]
    \begin{center}
      \includegraphics[width=0.6\textwidth]{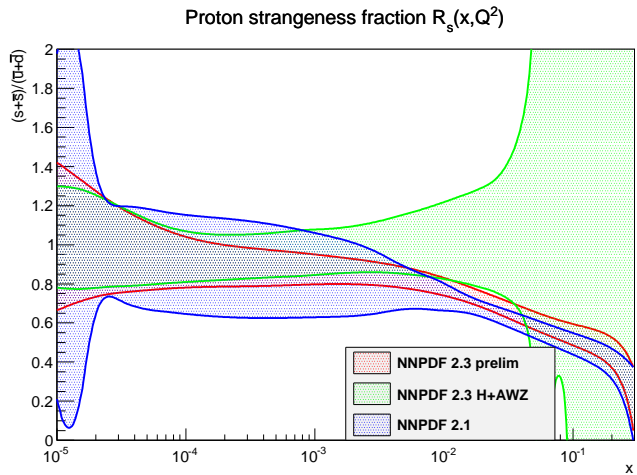}
    \end{center}
    \vskip-0.5cm
    \caption{\small Comparison of the proton strangeness fraction determined from fits to various datasets.}
    \label{fig:rs-xplot}
\end{figure}

A study by the ATLAS collaboration on the strange content of the proton,\cite{Aad:2012sb} based upon fits to the same dataset as NNPDF2.3 HERA+ATLASWZ suggest that the ratio of strange to non-strange PDFs may be underestimated by previous determinations from global fits. In Figure \ref{fig:rs-xplot} we examine the ratio of strange to non strange PDFs  for the NLO fits NNPDF2.3 prelim, NNPDF2.3 HERA+ATLASWZ, and NNPDF2.1. From this figure it is clear that the recent ATLAS W/Z measurements provide a valuable constraint, however at medium to large-$x$ the HERA and ATLAS data alone is insufficient to provide a precise determination of the strangeness.  

\begin{figure}[ht]
    \begin{center}
      \includegraphics[width=0.45\textwidth]{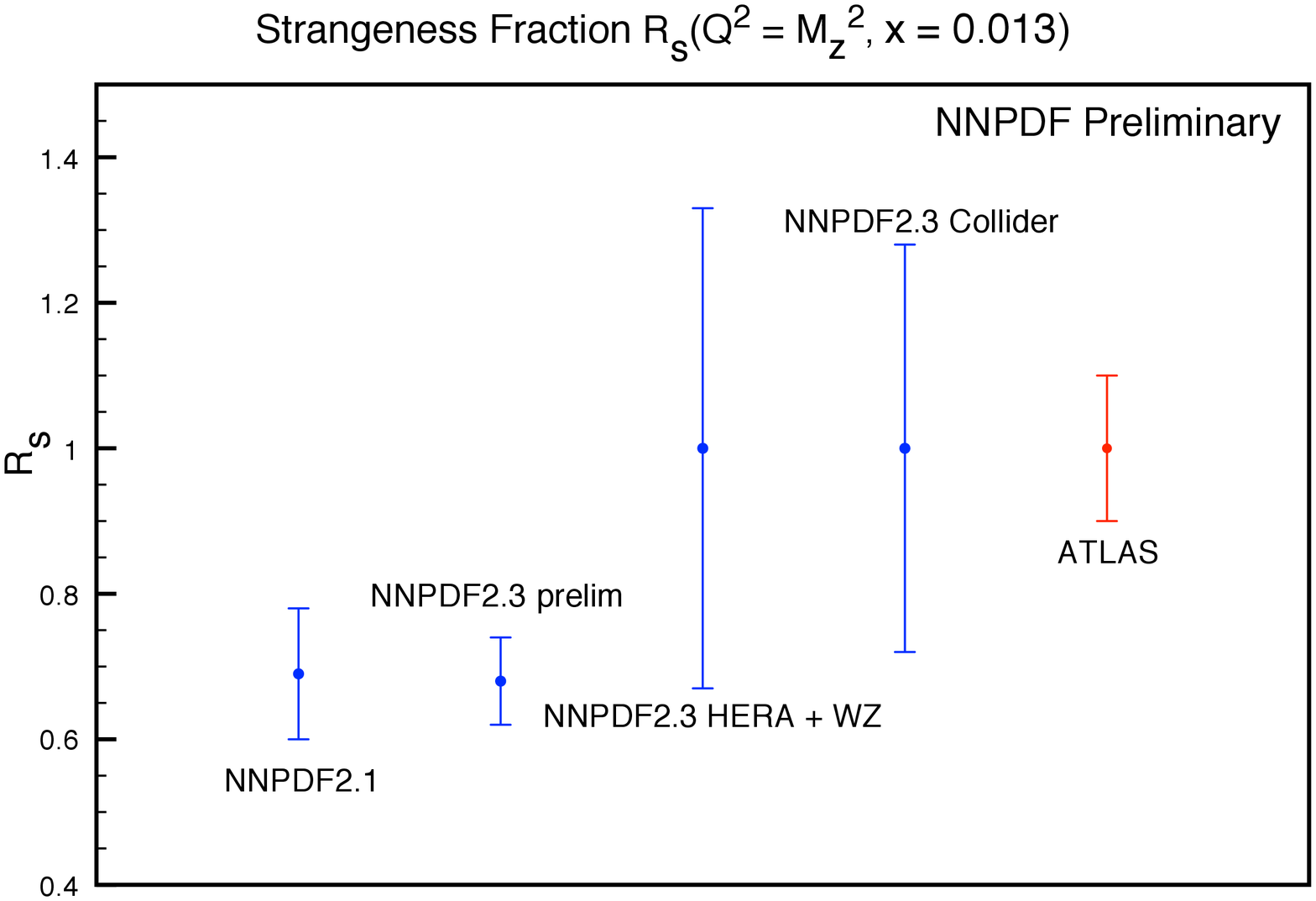}
           \includegraphics[width=0.45\textwidth]{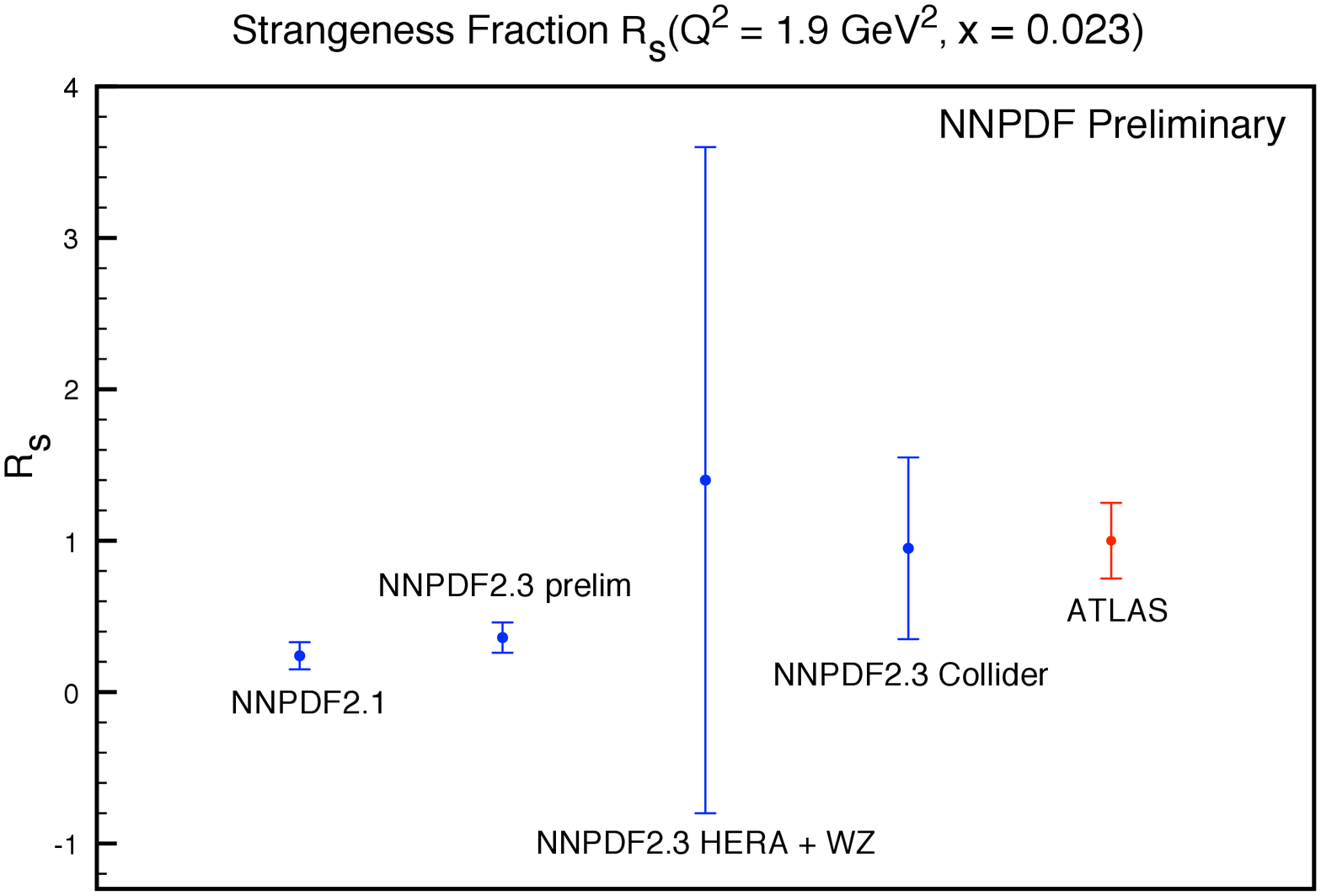}
    \end{center}
    \vskip-0.5cm
    \caption{\small Value of $R_s$ determined with different PDF sets at two choices of $(x,Q^2)$. }
    \label{fig:rs-comp}
\end{figure}

In Table 1 we compare the values obtained by the different sets at specific values of $(x,Q^2)$ and see a similar pattern. Fits to reduced datasets, such as the collider only
and HERA + ATLAS W/Z fits suggest a higher value of $R_s$, however the data provides little constraint and therefore the uncertainties are substantially larger than in the determinations provided by the global fits. The values all broadly agree within the large uncertainties of the HERA + ATLAS W/Z fit as shown in the comparison in Figure \ref{fig:rs-comp}.
We can therefore conclude that the collider data alone is not yet sufficiently constraining to provide a precise determination of the proton strangeness fraction, and that the uncertainty on $R_s$ in the ATLAS determination\cite{Aad:2012sb} has been underestimated.
\begin{table}[h]
\begin{center}
\begin{tabular}{|c|c|c|}
\hline
\centering PDF Set &  $R_s(0.013,M_z^2)$  & $R_s(0.023,1.9\mathrm{GeV}^2)$\\
\hline
\hline
NNPDF2.1   &  $0.61  \pm 0.09$ &   $0.24  \pm 0.09$ \\ 
NNPDF2.3  preliminary &	$0.68 \pm0.06$ & $0.36 \pm0.10$\\
\hline\hline
NNPDF2.3 HERA+ATLAS WZ only 	&	$1.00\pm0.33$&	$1.40\pm2.20$\\
NNPDF2.3 Collider only 		&$1.00\pm0.28$&$0.95\pm0.60$\\
\hline
\end{tabular}
\end{center}
\label{tab1}
\caption{Table of $R_s$ values determined from several PDF fits and at two choices of $(x,Q^2)$}
\end{table}
\section*{Acknowledgments}
The author would like to thank the organisers of the Rencontres de Moriond 2012 for the opportunity to present this work. This contribution made use of resources provided by the Edinburgh Compute and Data Facility (ECDF) (http://www.ecdf.ed.ac.uk/). The ECDF is partially supported by the eDIKT initiative (http://www.edikt.org.uk).

\end{document}